\documentclass[11pt,preprint]{aastex}

\usepackage{natbib}
\newcommand{\Fermi}{{\textit{Fermi}}}
\newcommand{\Swift}{{\textit{Swift}}}
\newcommand{\Epeak}{{E_{\rm peak}}}
\newcommand{\pfrac}[2]{\left(\frac{#1}{#2}\right)}
\newcommand{\Time}[1]{{$T_0 + #1$\,s}}
\newcommand{\e}{\epsilon}
\newcommand{\interval}[2]{{$T_0 + #1$\,s to $T_0 + #2$\,s}}

\renewcommand{\min}{{\rm min}}
\renewcommand{\max}{{\rm max}}

\begin{document}

\title{\Fermi\ Observations of GRB~090510: A Short Hard Gamma-Ray Burst
  with an Additional, Hard Power-Law Component from 10 keV to GeV Energies }

\author{
M.~Ackermann\altaffilmark{1}, 
K.~Asano\altaffilmark{2}, 
W.~B.~Atwood\altaffilmark{3}, 
M.~Axelsson\altaffilmark{4,5,6}, 
L.~Baldini\altaffilmark{7}, 
J.~Ballet\altaffilmark{8}, 
G.~Barbiellini\altaffilmark{9,10}, 
M.~G.~Baring\altaffilmark{11}, 
D.~Bastieri\altaffilmark{12,13}, 
K.~Bechtol\altaffilmark{1}
% ...
\\
~\\
Contact Authors:\\
  James Chiang (jchiang@slac.stanford.edu), 
  Jonathon Granot (j.granot@herts.ac.uk),\\
  Sylvain Guiriec (sylvain.guiriec@lpta.in2p3.fr),
  Masanori Ohno (ohno@astro.isas.jaxa.jp)}

\altaffiltext{1}{W. W. Hansen Experimental Physics Laboratory, Kavli Institute for Particle Astrophysics and Cosmology, Department of Physics and SLAC National Accelerator Laboratory, Stanford University, Stanford, CA 94305, USA}
\altaffiltext{2}{Interactive Research Center of Science, Tokyo Institute of Technology, Meguro City, Tokyo 152-8551, Japan}
\altaffiltext{3}{Santa Cruz Institute for Particle Physics, Department of Physics and Department of Astronomy and Astrophysics, University of California at Santa Cruz, Santa Cruz, CA 95064, USA}
\altaffiltext{4}{Department of Astronomy, Stockholm University, SE-106 91 Stockholm, Sweden}
\altaffiltext{5}{Lund Observatory, SE-221 00 Lund, Sweden}
\altaffiltext{6}{The Oskar Klein Centre for Cosmoparticle Physics, AlbaNova, SE-106 91 Stockholm, Sweden}
\altaffiltext{7}{Istituto Nazionale di Fisica Nucleare, Sezione di Pisa, I-56127 Pisa, Italy}
\altaffiltext{8}{Laboratoire AIM, CEA-IRFU/CNRS/Universit\'e Paris Diderot, Service d'Astrophysique, CEA Saclay, 91191 Gif sur Yvette, France}
\altaffiltext{9}{Istituto Nazionale di Fisica Nucleare, Sezione di Trieste, I-34127 Trieste, Italy}
\altaffiltext{10}{Dipartimento di Fisica, Universit\`a di Trieste, I-34127 Trieste, Italy}
\altaffiltext{11}{Rice University, Department of Physics and Astronomy, MS-108, P. O. Box 1892, Houston, TX 77251, USA}
\altaffiltext{12}{Istituto Nazionale di Fisica Nucleare, Sezione di Padova, I-35131 Padova, Italy}
\altaffiltext{13}{Dipartimento di Fisica ``G. Galilei", Universit\`a di Padova, I-35131 Padova, Italy}
% ...


\pagebreak

\begin{abstract}
We present detailed observations of the bright short-hard gamma-ray
burst GRB~090510 made with the Gamma-ray Burst Monitor (GBM) and Large
Area Telescope (LAT) on board the \Fermi\ observatory.  GRB~090510 is
the first burst detected by the LAT that shows strong evidence for a
deviation from a Band spectral fitting function during the prompt
emission phase.  The time-integrated spectrum is fit by the sum of a
Band function with $\Epeak = 3.9\pm 0.3$\,MeV, which is the highest
yet measured, and a hard power-law component with photon index
$-1.62\pm 0.03$ that dominates the emission below $\approx$\,20\,keV
and above $\approx$\,100\,MeV.  The onset of the high-energy spectral
component appears to be delayed by $\sim$\,0.1\,s with respect to the
onset of a component well fit with a single Band function. A faint GBM
pulse and a LAT photon are detected 0.5\,s before the main pulse.
During the prompt phase, the LAT detected a photon with energy
$30.5^{+5.8}_{-2.6}$ GeV, the highest ever measured from a short GRB.
Observation of this photon sets a minimum bulk outflow Lorentz factor,
$\Gamma\ga$\,1200, using simple $\gamma\gamma$ opacity arguments for
this GRB at redshift $z = 0.903$ and a variability time scale on the
order of tens of ms for the $\approx$\,100\,keV--few MeV flux.
Stricter high confidence estimates imply $\Gamma \ga 1000$ and still
require that the outflows powering short GRBs are at least as highly
relativistic as those of long duration GRBs.  Implications of the
temporal behavior and power-law shape of the additional component on
synchrotron/synchrotron self-Compton (SSC), external-shock
synchrotron, and hadronic models are considered.
\end{abstract}


\section{Introduction}

A difficulty in trying to understand gamma-ray bursts is that, at
least in terms of the temporal structure of their emission, all GRBs
differ.  When the overall time scales of the emission are considered,
however, a pattern does emerge.  The durations of the $\sim
100$\,keV--MeV emission from GRBs form a bimodal distribution and
hence are divided into two classes, namely the short and long duration
bursts. The short bursts formally have durations $< 2\,$s with typical
values around $\sim 0.2$\,s, whereas the long bursts have a
distribution that peaks around $\sim 30$\,s with a tail extending to
several hundreds of seconds \citep{Kouveliotou93}.

Although the physics of their gamma-ray emission is not well
understood, these two classes of bursts likely originate from distinct
types of progenitor systems \citep{Lee07, Woosley06}. Long duration
bursts are thought to be produced by the core collapse of massive
stars as evidenced by the direct association of several nearby GRBs
($z<0.3$) with SN Ib/c events \citep{Woosley06}.  Consistent with
this, the afterglow counterparts of long duration GRBs tend to lie in
star forming regions of low mass, irregular galaxies
\citep{Kocevski09}.  By contrast, short duration GRBs have been
associated with both early and late type host galaxies, in proportions
that reflect the underlying field galaxy distribution
\citep{Berger2009}. In the prevailing model for short bursts, they are
produced in merger events of a compact binary systems composed of two
neutron stars or a neutron star/black hole pair and so would tend to
originate from older stellar populations.

With the launch and successful operation of the \Fermi\ Gamma-ray
Space Telescope, a wider observational window has been opened through
which a greater understanding of GRBs may be obtained.  The Large Area
Telescope aboard \Fermi\ provides significantly greater energy
coverage (20 MeV to $>$\,300\,GeV), field-of-view ($2.4$\,sr) and
effective area ($8000$\,cm$^2$ at 1\,GeV) than its predecessor EGRET
\citep{2009ApJ...697.1071A}.  Owing to its substantially lower
deadtime (26\,$\mu$s vs 100\,ms for EGRET), the LAT can probe the
temporal structure of even the shortest GRBs.  In addition, the LAT
can localize GRBs with sufficiently high precision to enable follow-up
observations by \Swift\ and ground-based observatories; and at
energies $\ga$\,few\,GeV, the LAT can distinguish GRB photons from
background with little ambiguity.  The Gamma-ray Burst Monitor, the
other science instrument on \Fermi, comprises an array of 12 NaI
scintillators and two BGO scintillators and can detect gamma-rays from
8\,keV to 40\,MeV over the full unocculted sky. The combined
capabilities of the LAT and GBM enable \Fermi\ to measure the spectral
parameters of GRBs over seven decades in energy.  For sufficiently
bright bursts, time-resolved spectral analysis is possible over the
entire energy range.


The \Fermi\ observations of the short gamma-ray burst, GRB~090510,
take full advantage of the GBM and LAT capabilities.  The LAT emission
shows temporal structure on time scales as short as 20\,ms. In
addition to the usual Band function component \citep{Band:93},
spectral fits reveal a hard power-law component emerging in the LAT
band 0.1\,s after the onset of the main prompt emission in the GBM
band.  Moreover, a $\approx 0.2$\,s delay is observed between the
brightening of the $\approx 200$ MeV--GeV emission with respect to the
strong count increases in the NaI and BGO. These behaviors present
severe challenges for emission models of GRBs.

A photon with energy $30.5_{-2.6}^{+5.8}$\,GeV was detected by the LAT
0.829\,s after the GBM trigger.  This event arrived during the prompt
phase and is temporally coincident with a sharp feature in the GBM and
LAT light curves.  Given this energy, the temporal structure of the
burst light curve and the known distance to the burst,
\citet{Nature..GRB090510} have used this photon and its arrival time
to set limits on a possible linear energy dependence of the
propagation speed of photons due to Lorentz-invariance violation that
would require a quantum-gravity mass scale significantly above the
Planck mass.  Similar to several of the long bursts seen by
the LAT, GRB~090510 shows a high energy extended emission component
that is detected by the LAT as late as 200\,s after the GBM
trigger. In the context of GRB outflow models, the properties of this
GeV emission and the optical and X-ray afterglow observations by
\Swift\ place significant constraints on possible internal and
external shock models for the late time emission of this source
\citep{2009arXiv0909.0016G,2010ApJ...709L.146D,2009arXiv0911.4453C,2009ApJ...706L..33G,2009MNRAS.400L..75K}.

In this paper, we report on and discuss the GBM and LAT observations
of GRB~090510 during the prompt emission phase.  In section~2, we give
the basic observational details.  In section~3, we present a timing
analysis, including discussion of its designation as a short-hard GRB
and a cross-correlation analysis between various energy bands to
characterize any energy-dependent temporal lags.  In section~4, we
perform spectral analyses of both the time-integrated and the
time-resolved data; and we demonstrate the significance of the
additional power-law component and how the spectra evolves over the
course of the burst.  In section~5, we derive a lower limit on the
bulk Lorentz factor given the variability time scales and observations
of the highest energy photons, describe the constraints imposed on
leptonic, hadronic and other models of the prompt emission, and
discuss implications of the detection of the 30.5\,GeV photon for
models of the extragalactic background light (EBL).  Finally, we
summarize our results in section~6.

\section{Basic Observations}

On 2009 May 10 at 00:22:59.97 UT (hereafter $T_0$), both the GBM and
LAT instruments triggered on GRB~090510 (trigger 263607781). During
the first second after the trigger, the LAT detected 62 ``transient
class'' events with energies $>$\,100\,MeV and within 10$^\circ$ of
the \Swift\ UVOT position \citep{GCN..9342} and 12 events above
1\,GeV.  In the first minute post-trigger, the LAT detected 191 events
above 100\,MeV and 30 above 1\,GeV.  From spectral fits to the
time-integrated emission over the time range \interval{0.5}{1.0} (see
section~\ref{sec:spectral analysis}), the fluence of the burst is
$(5.03 \pm 0.25) \times 10^{-5}$\,erg\,cm$^{-2}$ in the 10\,keV to
30\,GeV band.  In the 15--150\,keV band, the fluence is $(4.08 \pm
0.07) \times 10^{-7}$\,erg\,cm$^{-2}$.

Other detections of the prompt emission were made by
\Swift\ \citep{GCN..9331}, AGILE \citep{GCN..9343}, Konus-Wind
\citep{GCN..9344}, Suzaku WAM \citep{GCN..9355}, and INTEGRAL-ACS.
The \Swift\ UVOT instrument measured the position of the optical
afterglow counterpart to be R.A.(J2000), Dec(J2000) = $22^{\rm
  h}14^{\rm m}12\fs5$, $-$26\arcdeg 34\arcmin 59.2\arcsec
\citep{GCN..9342}.  Follow-up optical spectroscopy taken 3.5 days
later by VLT/FORS2 measured [OII] and H$\beta$ emission lines at a
common redshift of $z = 0.903 \pm 0.001$ \citep{2010arXiv1003.3885M}.  Using
standard cosmological parameters ($[\Omega_\Lambda, \Omega_{\rm M}, h]
= [0.73, 0.27, 0.71]$), this corresponds to a luminosity distance of
$d_L = 1.80 \times 10^{28}$\,cm and implies an apparent isotropic
energy of $E_{\rm iso} = (1.08 \pm 0.06) \times 10^{53}$\,erg (10\,keV--30\,GeV, \interval{0.5}{1.0}).  The
optical afterglow was also detected by NOT \citep{GCN..9338} and GROND
\citep{GCN..9352}.  In the radio, however, only upper-limits were
obtained using the VLA \citep{GCN..9354}.

The host galaxy of GRB~090510 was identified as a late-type elliptical
or early-type spiral galaxy \citep{GCN..9353}, in contrast to the dwarf
irregular, star-forming galaxies that have been observed to harbor
long duration GRBs. This is consistent with the diverse types of hosts
identified with short GRBs \citep[e.g.,][]{2007PhR...442..166N, Berger2009}.

\section{Timing Analysis}

\begin{figure}
  \centering
  \includegraphics[scale=0.7]{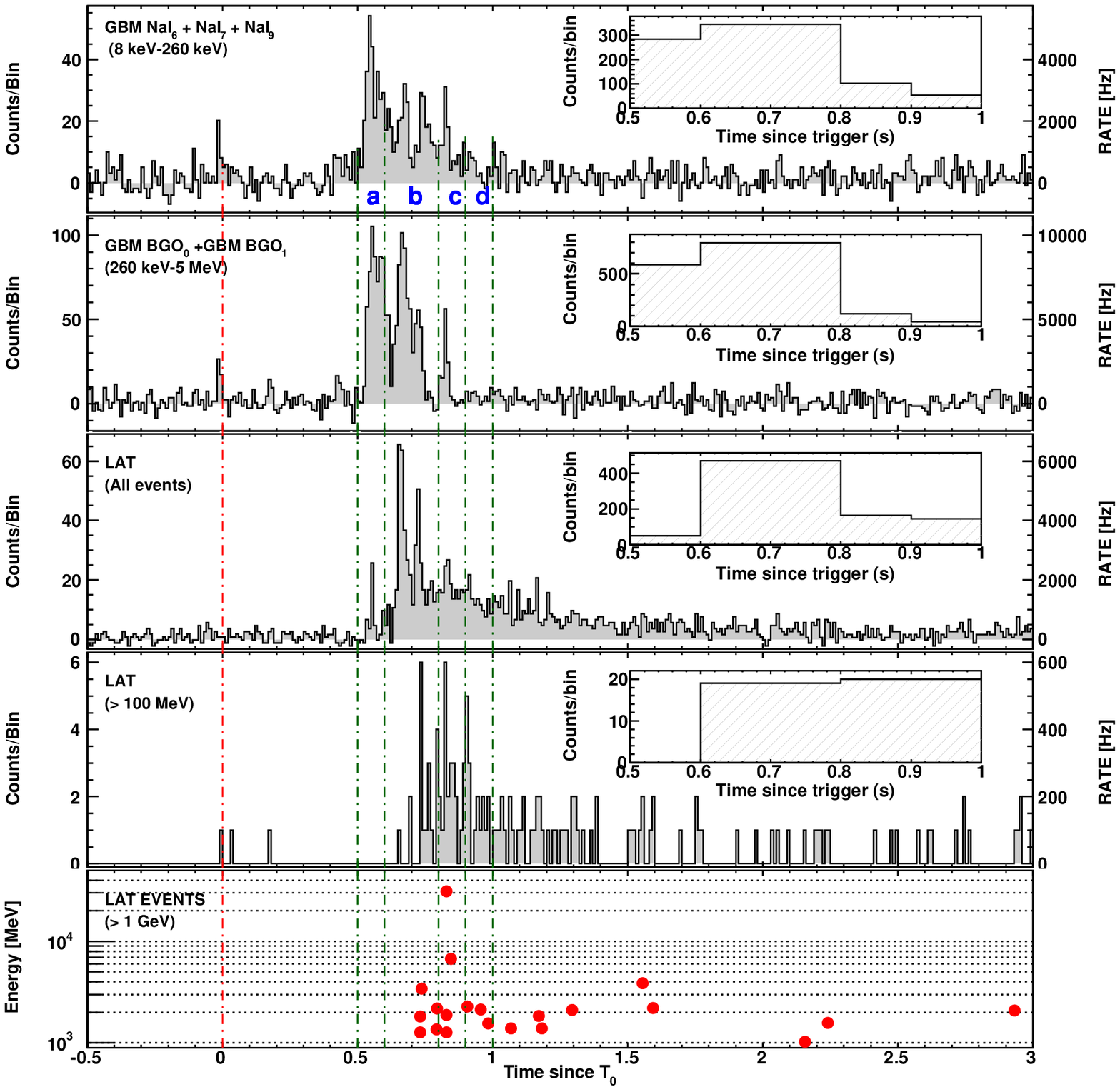}
  \caption{In the top four panels, the GBM and LAT light curves are
    shown, from lowest to highest energies. The bin size for all light
    curves is 0.01\,s.  In the bottom panel, the individual photon
    energies as a function of time are plotted.  The red vertical
    dot-dashed line indicates the trigger time, and the black vertical
    dot-dashed lines indicate the boundaries used for the
    time-resolved spectral analysis.  These intervals are labeled a,
    b, c, and d and correspond to \interval{0.5}{0.6},
    \interval{0.6}{0.8}, \interval{0.8}{0.9}, and \interval{0.9}{1.},
    respectively.  The insets in the top four plots show the counts
    per bin within those time intervals.  A previous version of this
    figure appeared in \citet{Nature..GRB090510}.}
  \label{fig:light curves}
\end{figure}

In Figure~\ref{fig:light curves}, we plot the light curves from the
different detectors in various energy bands.  The upper two panels
contain the data from the GBM NaI and BGO detectors in energy ranges
8--260\,keV and 0.26--5\,MeV, respectively.  Both of these light
curves show the precursor event at $T_0$ that caused the GBM trigger.
The main part of the emission in both of these light curves starts
approximately at \Time{0.5}, though both light curves have a small but
significant feature at \Time{0.4}.  The third panel displays all of
the LAT events that passed the on-board gamma-filter and have at least one reconstructed track, and the fourth
panel shows the light curve for the transient class selection at
energies $>$\,100\,MeV.  In the bottom panel, the measured photon
energies $>$\,1\,GeV are plotted versus arrival time.  The three LAT
events in the $>$\,100\,MeV light curve with arrival times in the
interval \interval{0}{0.2} are very likely burst photons.  Using the
count rate in the LAT during the 200\,s directly preceding $T_0$ as a
measure of the background rate, the probability that these three
photons would arise by accident is $1.2\times 10^{-6}$.

\begin{figure}
  \centering
  \includegraphics[scale=0.5]{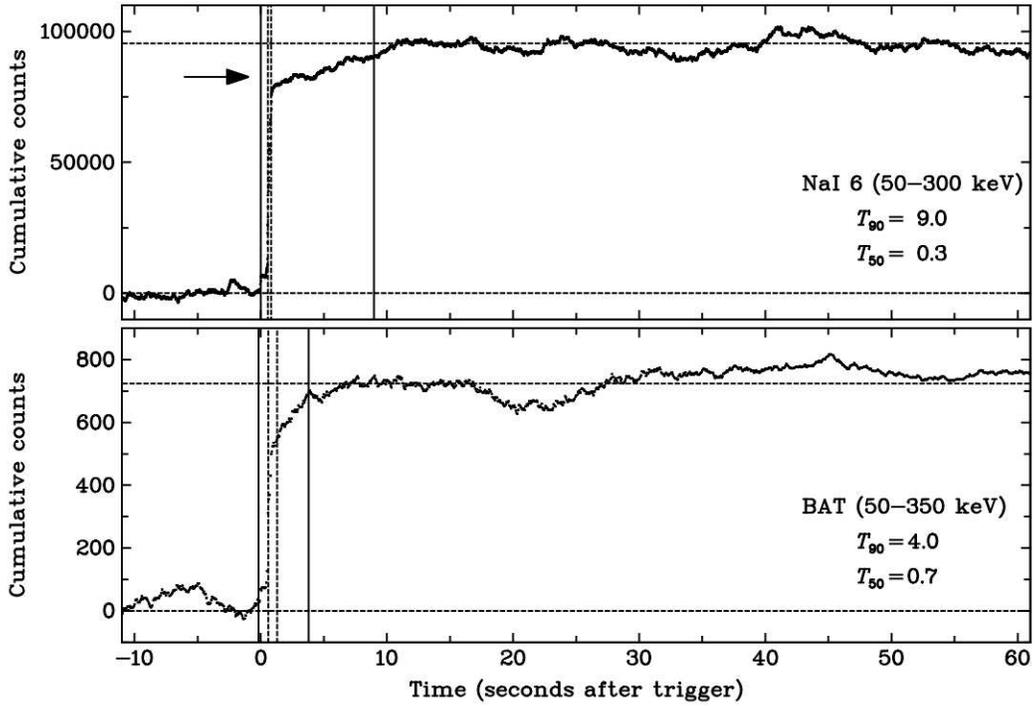}
  \caption{GRB~090510 $T_{90}$ and $T_{50}$ estimates using the
    GBM/NaI~6 (upper panel) and \Swift/BAT (lower panel) detectors.
    The horizontal dashed lines indicate the 0\% and 100\% cumulative
    plateau levels, and the vertical solid and dashed lines indicate
    the $T_{90}$ and $T_{50}$ interval boundaries, respectively. The
    arrow in the upper panel indicates the lowest alternative 100\%
    cumulative plateau level, which results in $T_{90} = 1.0$\,s.}
  \label{fig:durations}
\end{figure}

Classification of GRBs as short versus long bursts is based on the
$T_{90}$ or $T_{50}$ durations. The $T_{90}$ ($T_{50}$) duration is
defined as the time between accumulating 5\% and 95\% (25\% and 75\%)
of the counts associated with the GRB.  For the $T_{90}$ and $T_{50}$
durations of GRB~090510, we have applied the technique described in
\citet{Koshut1996} to determine these quantities using data from
multiple instruments (see Table~\ref{tab:durations}).  The cause of
the wide range in $T_{90}$ durations that we find is illustrated in
Figure~\ref{fig:durations}, where we have plotted the cumulative
counts for two detectors, GBM/NaI6 and \Swift/BAT, integrated over the
energy ranges 50--300\,keV and 50--350\,keV, respectively.  Ideally,
the selection of the 0\% and the 100\% plateaus, which designate the
onset and the end of the burst data accumulation, is unambiguous.
However, when large background variations exist, the plateau selection
is not unique as one can see in the upper panel of
Figure~\ref{fig:durations}. For the GBM data, the most conservative
selection provides $T_{90}=9.0$\,s, while alternative detector
selections give conservative lower values, down to 1.5\,s (see
Table~\ref{tab:durations}) . The \Swift/BAT data (lower panel) have a
much lower background and allow for a reasonably robust setting of the
100\% level and yield $T_{90}=4.0$\,s. Similar plots have been
obtained for other GBM detectors, as well as for the INTEGRAL-SPI and
the Suzaku-WAM instruments.  We find a narrower range of $T_{50}$
values, compared to the $T_{90}$ durations for the same dataset. The
latter are very sensitive to 5\% background variations, while the
former are more robust.

A common feature of long duration GRBs is the trend for their
higher-energy photons (below $\sim 1$\,MeV) to arrive before the
lower-energy ones \citep{Norris:2000}.  By contrast, short duration
GRBs typically show no evidence for lags between different energy
ranges below 1\,MeV \citep{Norris:06}.  Therefore, since the $T_{90}$
estimates in Table~\ref{tab:durations} span the formal 2\,s divide
between short and long duration bursts, the presence or absence of the
``hard-to-soft'' evolution seen in long bursts can help determine
whether GRB 090510 should be classified as a short burst.

\newcommand{\ccf}{\mbox{CCF}}

\begin{figure}
  \centering
  \includegraphics[scale=0.8]{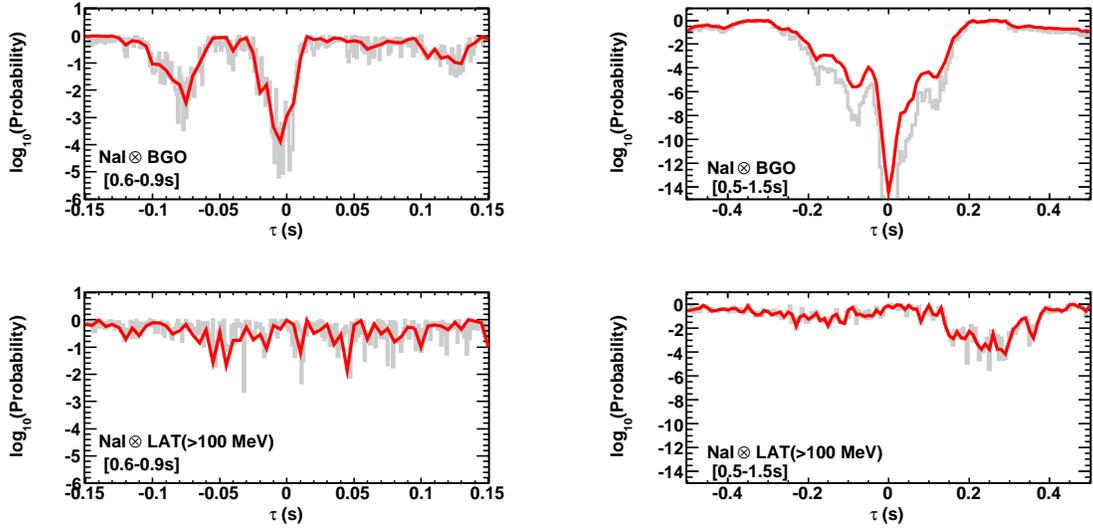}
  \caption{Cross-correlation function probability as a function of the
    lag $\tau$ for the specified light curve to follow the NaI light
    curve.  The time intervals considered are indicated in the
    square brackets in each panel.  The red curves correspond to time
    bins of size 0.01\,s, and the gray curves correspond to 0.005\,s.
    The fact that the probability curves are mostly independent of the
    bin size indicates that we are resolving all of the important
    features of the underlying variability.}
  \label{fig:ccfs}
\end{figure}

In order to estimate the energy-dependent lags for GRB 090510, we
compute the cross-correlation function (\ccf) between light curves for
different energy bands among the various detectors.  For these data,
we define the \ccf\ as a function of the lag $\tau$ using
\begin{equation}
\ccf(\tau) = \frac{\sum_{i=i_1}^{i_2}[f(t_{i})-\bar{f}][g(t_{i}+\tau)-\bar{g})]}
                  {\sqrt{\sum_{i=i_1}^{i_2}(f(t_{i})-\bar{f})^2}
                   \sqrt{\sum_{i=i_1}^{i_2}(g(t_{i}+\tau)-\bar{g})^2}}.
\label{eq:ccf}
\end{equation}
Here $f(t_i)$ and $g(t_i)$ represent the content of the light curves
we are comparing for bin $i$ at time $t_i = i\,dt$ and bin size $dt$;
$\bar{f}$ and $\bar{g}$ are the mean values of the two light curves
evaluated over the time interval considered, $(t_{i_1}, t_{i_2})$; and
the number of bins in each light curve is $N = i_2 - i_1 + 1$.  At a given
value of $\tau$, equation~(\ref{eq:ccf}) is simply the linear
correlation coefficient known as Pearson's $r$.  If $N$ is large, then
$r$ approximately follows a normal distribution with zero mean and
standard deviation $1/\sqrt{N}$.  In this case, the probability that a
value greater than $|r| = |\ccf(\tau)|$ can arise by chance between
two uncorrelated light curves is given by
\begin{equation}
P(|r|) = 1 - \mbox{erf}(|r|\sqrt{N/2}),
\label{eq:ccf_probability}
\end{equation}
where $\mbox{erf}$ is the error function. Since the amplitude of the
variations of the \ccf\ are a strong function of the bin size, we
instead plot the probability $P(\tau) = P(|\ccf(\tau)|)$ as a function
of the lag $\tau$.  As we show in Figure~\ref{fig:ccfs}, $P(\tau)$ is
largely insensitive to the bin size and yields probabilities that we
can easily interpret.

The emission in the LAT band is suppressed during \interval{0.5}{0.6}
relative to the emission in the GBM band, and that suppression can
dominate the \ccf\ probability curves and appear as a relative lag.
Since we also wish to determine if the individual pulses during the main
part of the burst are correlated (i.e., where most of the ``spiky''
structure is seen), we consider two time intervals for our
\ccf\ studies: \interval{0.5}{1.5} and \interval{0.6}{0.9}.  The
former interval includes essentially all of the prompt phase emission,
whereas the latter will be more useful for evaluating correlations in
the spiky structure.

In the two leftmost plots in Figure~\ref{fig:ccfs}, we show the \ccf\
probabilities for the \interval{0.6}{0.9} interval.  The upper plot
compares the NaI data with the BGO data.  The most significant
correlation is at zero lag.  Since the NaI data cover energies
8--260\,keV and the BGO data cover 260\,keV--5\,MeV, this result shows
that there is no evidence for an energy-dependent lag in the main Band
function component of the prompt emission.  This is consistent with
GRB~090510 being a short burst.  There is also a secondary minimum at
$\tau = -0.075$\,s, and this value for the lag corresponds to the rough
separation between the pulses in the NaI and first two pulses in the
BGO light curves and a shift of the BGO light curves by this amount to
later times.  In the lower left plot, no significant correlation at
any lag is seen between the NaI light curve and the LAT data above
$100$\,MeV; and we have performed Monte Carlo simulations and confirmed
that the lack of a correlation is not simply due to the relatively low 
statistics in the LAT data.
If indeed there is no such intrinsic temporal correlation, this would
strongly suggest a different emission region from the Band spectral
component, and no coupling between the observed emission from these
two regions, e.g., via the external Compton mechanism: soft photons
from the Band component upscattered to high energies by relativistic
electrons in a different region.
In the two rightmost plots of Figure~\ref{fig:ccfs}, the
results for the \interval{0.5}{1.5} interval are shown.  The strong
correlation at zero lag for the NaI versus BGO data is confirmed.  In
the NaI versus LAT data, the observed lag arises from the delayed
onset of the $>$\,100\,MeV emission by $\sim 0.2$--$0.3$\,s.

In addition to the lack of time lags for lower energy photons, short
GRBs are also known to be substantially harder than long GRBs.  For
BATSE bursts, \citet{Kouveliotou93} found a strong anticorrelation
between the $T_{90}$ duration and the hardness ratio of the burst. As
we will show in the next section, GRB~090510 has the highest peak
energy ever measured for any kind of burst and
is undoubtedly one of the hardest GRBs seen.  Given this result, the
range of $T_{90}$ and $T_{50}$ durations we find, and the lack of any
temporal lags for lower energy photons, we can safely consider
GRB~090510 to be a short duration GRB.

Finally, in Table~\ref{tab:Gamma_min_results}, we give the shortest
variability timescales, $t_v$, for the \interval{0.6}{0.8} and
\interval{0.8}{0.9} intervals.  These timescales will be used to
compute the minimum bulk Lorentz factor implied by the observation of
the highest energy ($\ga$\,GeV) photons in those epochs
(Section~\ref{sec:discussion}).  The $t_v$ values are given by the
full-width at half-maximum of the shortest pulse measured in any of
the detectors for a given time interval.

\section{Spectral Analysis}
\label{sec:spectral analysis}

The spectral analysis included data from the most brightly illuminated
GBM/NaI detectors, 3, 6, 7, 8, \& 9, covering an energy range
8\,keV--1\,MeV, and from both GBM/BGO detectors, covering
200\,keV--40\,MeV.  Since we wish to fit spectra on short time scales,
we used the Time-Tagged Event (TTE) data with the energy overflow
channels removed.  The background for each GBM detector was found by
fitting the data in 30\,s time intervals that preceded and followed
the prompt emission component by $\sim\,2$\,s, using a polynomial
function, then extrapolating to the times during the burst.
Custom-made detector response files for the different GBM detectors
were created using the \Swift/UVOT location.  The LAT photon data were
extracted for energies 100\,MeV--200\,GeV and using an
energy-dependent 95\% PSF acceptance cone centered on the source,
where we have increased the acceptance cone radius by adding in
quadrature the uncertainty in the \Swift/UVOT location.  Appropriate
for short time scale ($\ll$\,1\,ks) analyses, we used the
``transient'' class event selection and the corresponding instrument
response functions known as P6\_V3\_TRANSIENT.  The response matrix
files were created using the {\tt gtrspgen} tool from ScienceTools
v9r15p2. The background during the burst was computed by averaging the
LAT background over several orbits of the spacecraft during epochs
when it had the same orbital position and pointing.  The joint
spectral fits were performed using the spectral analysis package RMFIT
(version 3.1).

In Table~\ref{tab:spectral fits}, we show the results of joint
spectral fits of the GBM and LAT data for the time-integrated and
time-resolved data, including the definitions of the intervals used
for spectral analysis. We define four intervals a, b, c, and d corresponding
to \interval{0.5}{0.6}, \interval{0.6}{0.8}, \interval{0.8}{0.9},
and \interval{0.9}{1.}, respectively.
%
The time-integrated spectrum from \Time{0.5} to \Time{1.0} shows a
deviation from the standard Band function and can be adequately fit
with the addition of a power-law spectral component,
\begin{equation}
  n(E) = A\left(\frac{E}{1\,\mbox{GeV}}\right)^\lambda,
\end{equation}
or a Comptonized spectral component,
\begin{equation}
  n(E) = A\left(\frac{E}{1\,\mbox{GeV}}\right)^\alpha
         \exp\left(-\frac{E(2 + \alpha)}{E_{\rm pk}}\right).
\end{equation}
The Band function consists of two power-laws that
are smoothly joined near the peak photon energy, $\Epeak$.  For our
analyses, we use the Band function in the form
\begin{equation}
\begin{array}{rclr}
n(E) &=& A \pfrac{E}{100\,{\rm keV}}^\alpha 
           \exp\left(-\frac{E(2+\alpha)}{\Epeak}\right) & E < E_c, 
           \nonumber\\
     &=& A \pfrac{(\alpha-\beta)\Epeak}{100\,{\rm keV}(2+\alpha)}^{\alpha-\beta}
           \exp(\beta - \alpha)\left(\frac{E}{100\,{\rm keV}}\right)^\beta
           & E \ge E_c
\end{array}
\label{eq:Band}
\end{equation}
where $E_c = (\alpha - \beta)\Epeak/(2 + \alpha)$ \citep{Band:93}.

\begin{figure}
  \centering
  \includegraphics[angle=90,scale=0.6,clip]{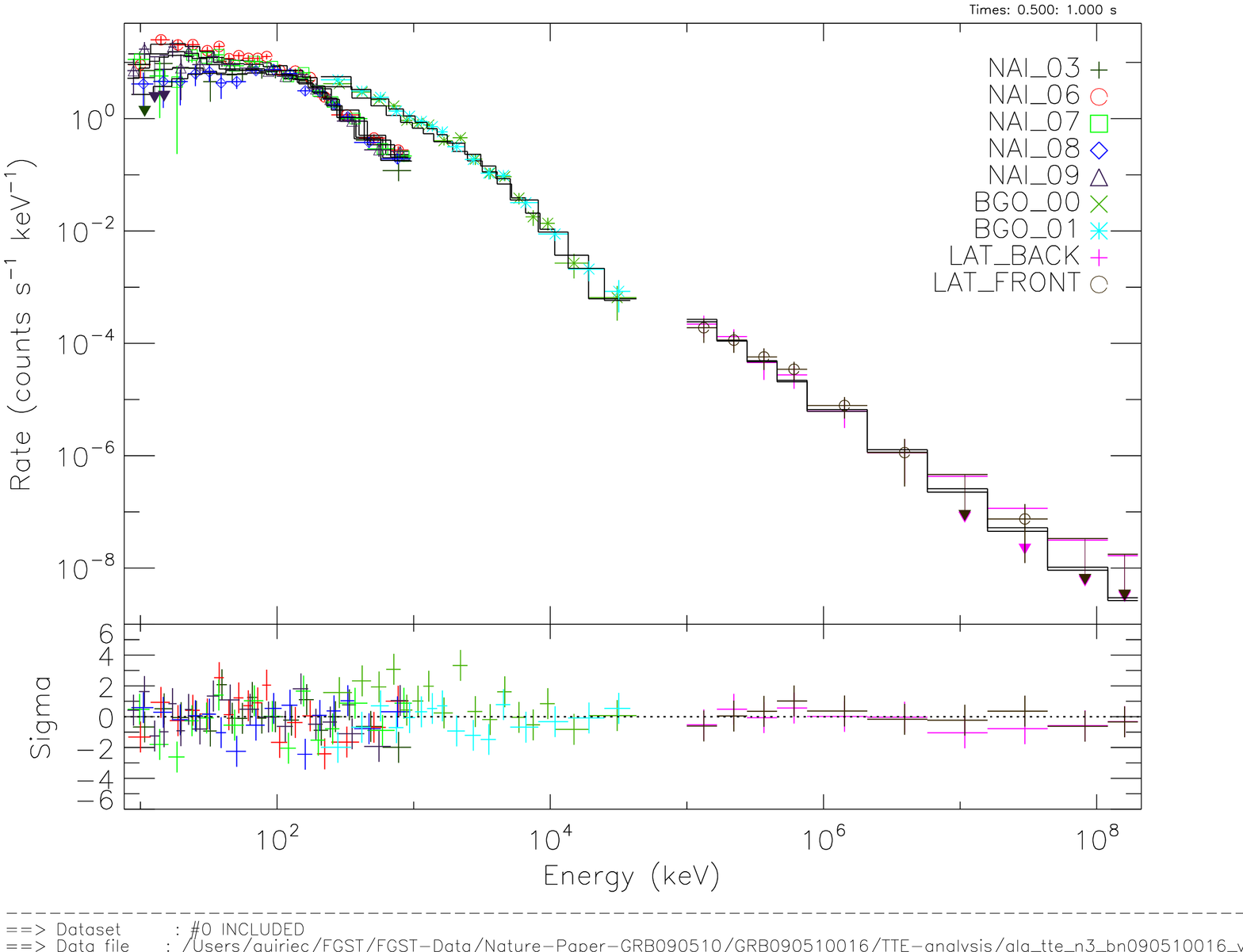}
  \caption{Counts spectrum for the time-integrated ($T_0 +
  0.5, T_0 + 1.0$\,s) data.  The Band + power-law model has been fit
  to these data. See Table~\ref{tab:spectral fits}.}
  \label{fig:counts spectra}
\end{figure}

For the time-integrated data, the peak energy of the Band component is
$\Epeak = 3.9 \pm 0.3$\,MeV.  This is the highest peak energy ever
measured in a GRB time-integrated spectrum.  The addition of the
power-law component with a photon index of $-1.62 \pm 0.03$
significantly improves the fit by more than 5\,$\sigma$ compared to a
single Band function.  This is the first short burst for which such a
hard power-law component has been measured.  The power-law component
appears to extrapolate to energies well below $\Epeak$ and dominates
the Band function emission below $\approx\,20$\,keV, similar to the
behavior seen in GRB~090902B \citep{Fermi_GRB090902B}.
Figure~\ref{fig:counts spectra} shows the counts spectrum of the
time-integrated data and the Band function + power-law fit.  In
Figure~\ref{fig:nuFnu spectra}, we plot this composite model as a $\nu
F_\nu$ spectrum and also plot the separate contributions for each
component.

For the time-resolved spectroscopy, we initially considered the data
partitioned into 0.1\,s time bins, starting at \Time{0.5}.  However,
for the analyses we present here, we have combined the data in the
\interval{0.6}{0.8} interval into a single bin in order to have
sufficient counts at energies $>$\,100\,MeV to constrain the fit of
the LAT data.  The parameters from the spectral fits to the selected
intervals are given in Table~\ref{tab:spectral fits}.  The Band
component undergoes substantial evolution over the course of the
prompt phase, starting out relatively soft with $\Epeak \approx
3$\,MeV, evolving to a very hard spectrum with $\Epeak \approx
5$\,MeV, accompanied by the appearance of the power-law component at
$>$\,100\,MeV, and then becoming softer again with $\Epeak \approx
2$\,MeV.  The extra power-law component hints at a similar sort of
soft-hard-soft evolution, but these spectral changes do not appear to
be commensurate with the Band component evolution (see
Figure~\ref{fig:nuFnu spectra}b).

\section{Discussion and Implications}
\label{sec:discussion}

The emergence of a distinct high-energy spectral component in the
prompt-phase spectrum of GRB~090510 establishes that a hard emission
component in addition to a Band component is found in the short hard
class of GRBs.  Hard power-law components are also found in three
long-duration GRBs, namely GRB~090902B \citep{Fermi_GRB090902B},
GRB~090926A \citep{Fermi..GRB090926A}, and GRB~941017
\citep{Gonzalez:03}.  In GRB~090510, the LAT data show that the
$\gamma$-ray flux above 100 MeV brightens $\approx 0.2$\,s after the
start of the bright phase of GBM emission.  This behavior is similar,
but on a shorter timescale, to the delayed onset in the long-duration
GRB~080916C \citep{2009Sci...323.1688A} and most other \Fermi\ GRBs,
including GRB~081024B, the first short GRB observed with the LAT 
\citep{2010ApJ...712..558A}.  Furthermore, GRB~090510 displays
$>$\,100\,MeV emission significantly extended beyond the duration of
the GBM flux \citep{2010ApJ...709L.146D}, as observed in other \Fermi\
GRBs and earlier from GRB 940217 with the EGRET instrument on the {\it
Compton Observatory} \citep{Hurley:94}.  These behaviors provide
important constraints for high-energy emission models and could help
answer whether the high-energy $\gamma$ rays have a leptonic or
hadronic origin.

Though the afterglow radiation in both long and short GRBs is probably
nonthermal synchrotron emission from an external shock
\citep{1998ApJ...497L..17S}, the situation is less clear in the prompt
and early afterglow phases when the GRB engine is most powerful.  This
radiation could be from the thermal photosphere made by the powerful
relativistic wind \citep{2002ApJ...578..812M,2007ApJ...664L...1P},
from magnetic reconnection in Poynting-flux dominated outflows
\citep{2003astro.ph.12347L}, or from nonthermal leptonic emissions
formed by internal or external shocks
\citep[e.g.,][]{2009MNRAS.400L..75K,2009arXiv0909.0016G,2009arXiv0911.4453C} 
in the relativistic jet of a GRB.

\citet{2010ApJ...709L.146D} describe and present models for the
\Swift\ and \Fermi\ observations of GRB~090510 during the afterglow
phase.  Here we consider the implications of the prompt phase and
early afterglow emission for GRB 090510.  After deriving the minimum
bulk Lorentz factor $\Gamma_\min$ and considering the various
uncertainties that enter into this calculation, we use the
observations to constrain leptonic synchrotron/SSC model and hadronic
models of short duration GRBs.  We do not discuss a thermal
photospheric interpretation for the \Fermi\ results on GRB~090510.
The photospheric interpretation overcomes the problem that the GBM
spectra are harder than expected below $\Epeak$ with the
simplest synchrotron emission model (which is only the case at
$\gtrsim 2\sigma$ in GRB~090510 during interval b; see
Table~\ref{tab:spectral fits}).  Even if it explains much of the GBM
emission, however, a different origin is needed for
the separate hard spectral component observed at LAT
energies.  The coincident narrow spikes between the LAT all events
and GBM lightcurves would not be easy to explain in a purely
photospheric scenario, though Compton-scattered photospheric emission
by internal shocked electrons could produce the coincident components
\citep{2010arXiv1002.2634T}.

\subsection{Lower Limit on the Bulk Lorentz Factor}

The use of $\gamma$-ray observations to constrain the bulk outflow
speed of highly variable and energetic $\gamma$-ray emission from GRBs
has been studied by many authors
\citep[e.g.,][]{1997ApJ...491..663B,2001ApJ...555..540L,2004ApJ...613.1072R}.
A detailed derivation involves an integration over the photon spectrum
to calculate the opacity of $\gamma$ rays emitted from sources with
idealized geometries \citep[see Supplementary Information
in][]{2009Sci...323.1688A}.  A $\delta$-function approximation for the
$\gamma\gamma$ opacity constraint gives values of $\Gamma_\min$
accurate to $\sim 10$\% whenever the target photon spectrum is softer
than $\nu F_\nu \propto \nu$.  In this case, $\gamma\gamma$ opacity
arguments for a $\gamma$-ray photon with energy $\e_1$, in $m_ec^2$
units, imply a minimum bulk Lorentz factor, defined by $\tau_{\gamma\gamma}(\e_1) =
1$, of \citep[e.g.][]{1995MNRAS.273..583D,2007PhR...442..166N},
\begin{equation}
\Gamma_{\rm min} \cong \left[ {\sigma_{\rm T} d_L^2 (1+z)^2 
f_{\hat \e}\e_1\over 4t_v m_ec^4}\right]^{1/6}\;\;,\;\;
\hat \e = {2\Gamma^2\over (1+z)^2\e_1}\;.
\label{eqs2}
\end{equation}
Here $f_\e $ is the $\nu F_\nu$ flux at photon energy $m_ec^2
\epsilon$, which is evaluated at $\e = \hat \epsilon$ due to 
the peaking of the $\gamma\gamma$ cross section near
threshold.  While the local value of the photon index
around $\hat\epsilon$ has some effect on the exact numerical
coefficient, this effect is small provided that the target photon index
is $< -1/2$. Because of the threshold condition used to relate the
high-energy photon and the target photons, the solution to
equation~(\ref{eqs2}) is iterative but quickly converges.  We use this
expression to estimate $\Gamma_\min$ from \Fermi\ observations of GRB
090510 for comparison with more accurate calculations.

For interval b, during which a 3.4\,GeV photon was detected, spectral
analysis of GBM and LAT data during this episode reveals distinct
Band-function and power-law components (Figure~\ref{fig:nuFnu spectra}). The
Band function has $\Epeak = 5.1$ MeV, $\alpha = -0.48$ and $\beta =
-3.09$ (Table~\ref{tab:spectral fits}). The combined Band plus
power-law fit reaches a peak $\nu F_\nu$ flux of $\approx 4\times
10^{-5}$ erg cm$^{-2}$ s$^{-1}$.  Writing the variability timescale
$t_v$(s)$=0.01 t_{-2}$ s for 10 ms variability timescale, and $f_{\hat
  \e} = 10^{-5} f_{-5}$ erg cm$^{-2}$ s$^{-1}$, then equation~(\ref{eqs2})
gives $\Gamma_\min \cong 1100(f_{-5}/t_{-2})^{1/6}\equiv 10^3\Gamma_3$
with $\epsilon_1 = 3400/0.511 \cong 6650$. The target photon energy $\hat \e \cong
2\Gamma_\min^2/(1+z)^2\e_1 \cong 110 (f_{-5}/t_{-2})^{1/3}$, or
$\approx 50$ MeV, corresponding to the Band $\beta$ branch of the
function.  Depending on whether the 3.4\,GeV photon is interacting
with the total emission or just the photons in the power law, then $f_{-5}
\approx 0.7$ or $f_{-5}\approx 0.1$, and $\Gamma_\min \cong 950$ or $\Gamma_\min \cong 720$, 
respectively. For interval c
from \interval{0.8}{0.9}, the same procedure with the 30.5\,GeV
photon gives $\Gamma_\min \cong 1370$ or 1060 for $f_{-5}
\approx 0.5$ or $f_{-5}\approx 0.1$ corresponding respectively to the 
combined Band plus power-law fit or the power-law component only.

The results of numerical integrations to determine $\Gamma_\min$ using
the more detailed expressions in \citet{2009Sci...323.1688A} are shown
in Table~\ref{tab:Gamma_min_results}.  As can be seen, the simple estimates given above are in
good agreement with the detailed calculation.  A number of issues
arise in the use of equation~(\ref{eqs2}) or the numerical
integrations that are important for assessing the value and
uncertainty in $\Gamma_\min$. These include the error incurred by the
uncertainties in source spectral fitting parameters, which properly
involves a covariance matrix to correlate uncertainties for different
parameters of the Band-function and power-law fit.  

For $E_{\rm max}$, we take the
highest energy photon associated with that pulse.
Table~\ref{tab:Gamma_min_results} presents the values for $t_v$,
$E_{\rm max}$, and $\Gamma_\min$ for time intervals
\interval{0.6}{0.8} and \interval{0.8}{0.9}, which are the only two
for which a distinct pulse width could be measured.  In interval
\interval{0.6}{0.8}, the Band + PL fit shown in
Table~\ref{tab:spectral fits} form the target photon spectrum, while
in interval \interval{0.8}{0.9}, we present results for both the Band
and Band + PL fits since each fits the data reasonably well. We also give
values for $\Gamma_\min$ assuming that only the hard power-law component
forms the target photon source. Here we assume that $t_v$ is the same as 
that measured from the BGO emission, which is primarily associated with photons
in the Band portion of the spectrum. If the variability timescale of the 
power-law emission is different than assumed, which would be compatible with the 
two components originating from different locations, then the minimum 
Doppler factor limit would change as indicated by equation~(\ref{eqs2}). 
Furthermore, for calculations of $\Gamma_\min$ we use the spectrum derived on 
0.2 s (time interval b) and 0.1 s (time interval c) timescales rather than on the 
shorter variability timescales during which the high-energy photons are measured. 
This is required for accurate spectral analysis, 
but could underestimate the flux (and therefore $\Gamma_\min$) 
during the bright narrow spikes, as can be seen from Fig. \ref{fig:light curves}.

The derivation of $\Gamma_\min$ depends crucially on the assumption
that the high-energy radiation and target photons are made in the same
emitting region.  Correlated variability between different wavebands
would support the cospatial assumption (see Figure~\ref{fig:light
  curves}), but no strong evidence for this behavior was found in the 
\ccf\ analysis described in Section~3.  
A conservative assumption would be to suppose that the high-energy
photon is part of the power-law component and that it can potentially
interact only with target photons that are part of the same power-law
emission component.
Even in cases where the target MeV photons are made at smaller radii 
than the high-energy photons, or in different regions within the 
Doppler cone of the emitting surface, spacetime overlap will add to 
opacity, so this should represent the most conservative assumption.
Further complicating the derivation of $\Gamma_\min$ is the assumed
emitting geometry and the temporal evolution of the radiating plasma.
For a blast-wave geometry, the precise value of $\Gamma_\min$ depends
on whether high-energy photons are produced throughout the ``shell''
or from the inner edge of the ``shell,'' and on the dynamical behavior of
the target photons \citep{Granot:08}.

Finally, a significant uncertainty on $\Gamma_\min$ can arise if
the photon with observed energy $E_{\rm max}$ is a random fluctuation
of the underlying true spectrum that corresponds to $\Gamma \lesssim
\Gamma_\min$ and $\tau_{\gamma\gamma}(E_{\rm max}) \gtrsim 1$.
The confidence we have on the value of $\Gamma$ depends on the
radiative transport and escape of $\gamma$ rays from the emitting
region. For interval c from \interval{0.8}{0.9}, $\Gamma_\min = 1218
\pm 61$ for the Band plus power-law fit
(Table~\ref{tab:Gamma_min_results}). Assuming that the intrinsic
spectrum extrapolates as a power law to high-energies, a likelihood
ratio test assuming an exponential escape probability gives
$\Gamma/\Gamma_\min = 0.96$, $0.88$, and $0.80$ and a spherical escape
probability gives $\Gamma/\Gamma_\min = 0.89$, $0.69$, and $0.49$ at the
1, 2, and $3\sigma$ confidence levels, respectively. 
The presence of two photons with energies between 1 and 2 GeV in 
interval b and a 7 GeV photon in interval c reduces the likelihood 
that the highest energy photon is a fluctuation and can be used to 
independently estimate $\Gamma_\min$, giving a value $\Gamma_\min
\gtrsim 1000$.

GRB 090510 is the second short GRB observed with LAT, after
GRB~081024B \citep{2010ApJ...712..558A}, but the first with a
redshift, which is required to derive $\Gamma_\min$. The value of
$\Gamma_\min \cong 1200$--1300 for GRB 090510 is comparable to, and
slightly larger than the values of $\Gamma_\min \cong 900$ and
$\Gamma_\min \cong 1000$ derived for GRB 080916C
\citep{2009Sci...323.1688A} and GRB 090902B \citep{Fermi_GRB090902B}
using corresponding $\gamma\gamma$ opacity arguments.  This has led to
suggestions that the GRBs with the most luminous LAT emission are
those with the largest bulk Lorentz factors
\citep{2009MNRAS.400L..75K,2009arXiv0909.0016G}.

\subsection{Models for the Prompt Radiation from GRB 090510}

In addition to the requirement of bulk outflow Lorentz factors $\Gamma
\gtrsim \Gamma_\min$, models for GRB 090510 should explain the
$\approx 0.2$\,s delay of the onset of the $\gtrsim 100$\,MeV emission
compared to the start of the main GBM emission at \Time{0.5}, the
appearance of a hard component, and the high-energy radiation
extending to $\approx$\,\Time{150}.

\subsubsection{Synchrotron/SSC Model}

A standard GRB model for the prompt phase assumes that the keV--MeV
emission is nonthermal synchrotron radiation from shock-accelerated
electrons \citep[e.g.,][]{1996PhRvL..76.3478T}.  This emission is
necessarily accompanied by SSC radiation.  

The SSC component is
stronger for a large
ratio of nonthermal electron to magnetic-field energy density, as
expressed by the condition $\epsilon_e \gg \epsilon_B$
\citep{2001ApJ...548..787S,Zhang:01}, and  also 
when the GRB has a lower bulk Lorentz factor for an external shock origin. 
Here $\epsilon_e$ and $
\epsilon_B$ are the fractions of shocked energy transferred to
nonthermal lepton and magnetic-field energy, respectively.  Lower bulk
Lorentz factors that give stronger SSC components result in greater
attenuation of high-energy $\gamma$ rays from $\gamma\gamma$
pair-production processes, as discussed in Section 5.1.  For generic
($E_{\rm iso} \sim 10^{52}$ erg, $\Gamma_0\sim 300, n\sim 1$
cm$^{-3}$) parameters expected in an external shock model, a
hardening of the LAT spectrum due to the deceleration of the blast
wave and the emergence of the SSC component in the LAT band was
expected to take place in the afterglow phase, but not in the prompt
phase \citep{Dermer:00b}.  With the larger initial Lorentz
factors $\Gamma_0\gtrsim 10^3$ implied by $\gamma\gamma$ arguments for
LAT GRBs and the earlier emergence of an external shock component, an
SSC component would persist after the decline of the synchrotron
component due to less scattering taking place in the Klein-Nishina
regime as the blast wave decelerates, and to the peak frequency of the
SSC flux decreasing from the TeV to the GeV range. The detailed
observations of GRB~090510 can help determine whether the hard
power-law component appearing at $\approx$\,\Time{0.7} can be
explained by SSC emission in the LAT waveband during the prompt phase.

\begin{figure}
  \centering
  \includegraphics[scale=0.85]{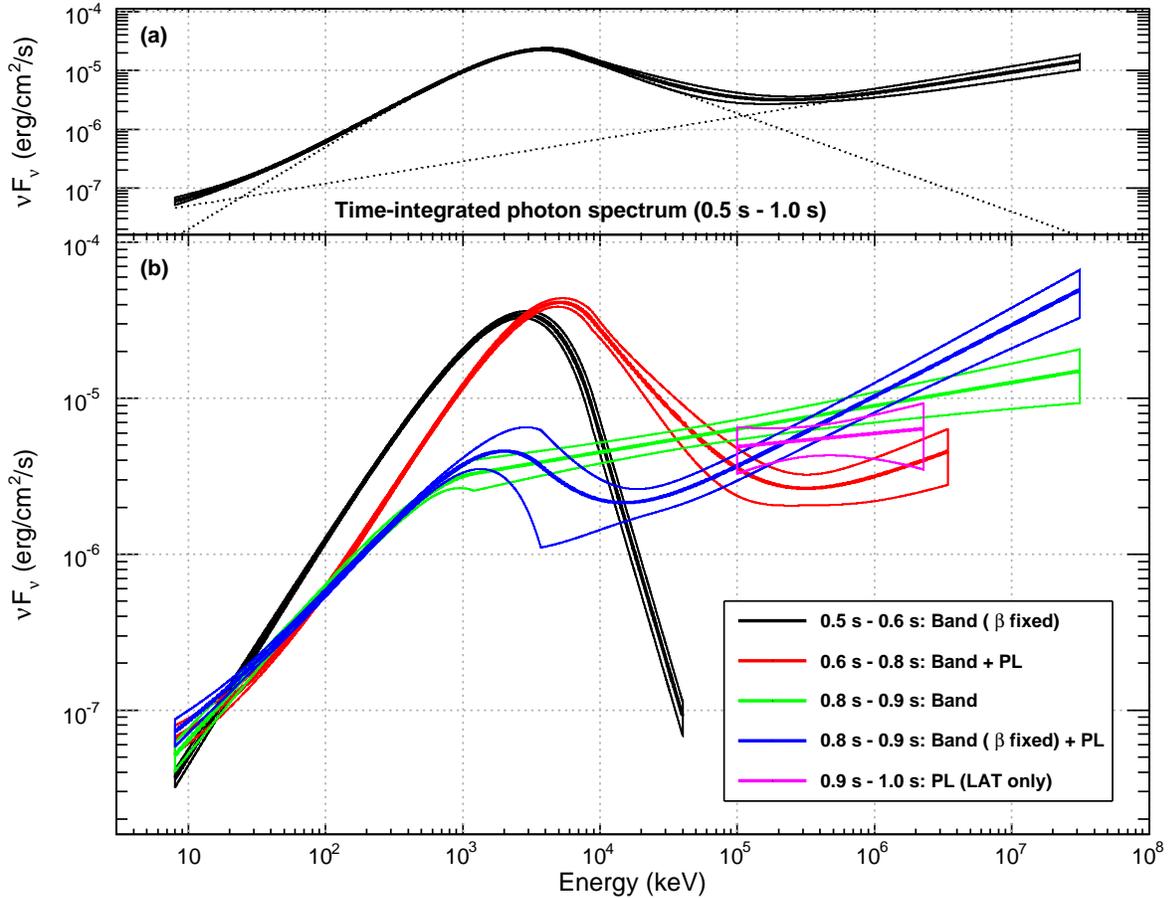}
  \caption{{\bf a.} The best-fit Band + power-law model for the
  time-integrated data plotted as a $\nu F_\nu$ spectrum. The two
  components are plotted separately and the sum is plotted as the
  heavy line.  The $\pm 1\,\sigma$ error contours derived from the
  errors on the fit parameters are also shown. {\bf b.}  The $\nu
  F_\nu$ model spectra (and $\pm 1\,\sigma$ error contours) plotted
  for the different time ranges given in Table~\ref{tab:spectral fits}.}
  \label{fig:nuFnu spectra}
\end{figure}

\begin{figure}
  \centering
  \includegraphics[scale=0.7]{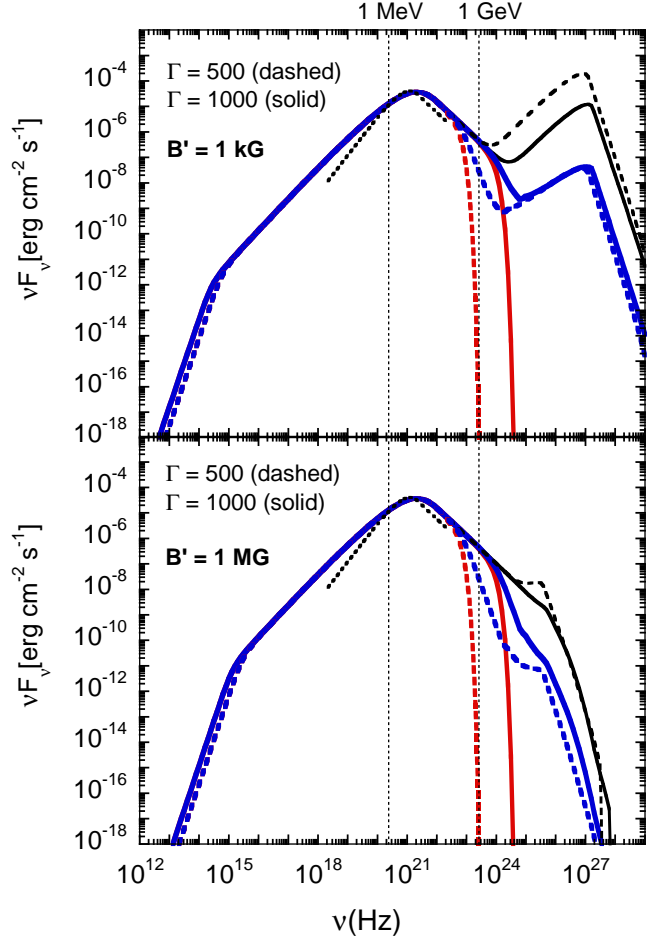}
  \caption{ Synchrotron/SSC model for interval b (\interval{0.6}{0.8})
    of GRB 090510. Dotted curves show the Band component during this
	time interval. The magnetic field in the top and bottom panels is
    1 kG and 1 MG, respectively. The dashed and solid curves show
    results for $\Gamma = 500$ and $\Gamma = 1000$, respectively.  The
    upper blue and lower red curves show received fluxes
    (without EBL corrections) assuming spherical and exponential
    $\gamma$-ray escape probabilities, respectively. }
  \label{fig:dermerSSC}
\end{figure}

Figure~\ref{fig:dermerSSC} shows results for a numerical model where
synchrotron peak energy, peak flux and variability time are made to
correspond to the observed values shown in Figure~\ref{fig:nuFnu spectra}.
This code employs a Compton kernel that accurately treats
Compton-scattering of relativistic electrons throughout the Thomson
and Klein-Nishina regimes, internal $\gamma\gamma$ opacity,
synchrotron self-absorption, radiative escape for $\gamma$ rays
described by exponential and spherical escape probabilities, and
second-order SSC.  The code does not include reprocessing of
internally-absorbed radiation or effects of attenuation of
$\gamma$-ray photons by the EBL (see Section 5.3).  Parameters
appropriate to the Band component in interval b (see
Figures~\ref{fig:light curves} and~\ref{fig:nuFnu spectra}) are used, as
shown by the dotted curves.  The parameters are $t_v = 14$\,ms,
$\Epeak = 5.10$\,MeV, $E_\max = 3.4$\,GeV, $\alpha = -0.48$, $\beta =
-3.09$, and peak synchrotron flux $\approx 4\times
10^{-5}$\,erg\,cm$^{-2}$\,s$^{-1}$ (see Figure~\ref{fig:nuFnu spectra} and
Tables~\ref{tab:spectral fits} and~\ref{tab:Gamma_min_results}).
Figure~\ref{fig:dermerSSC} shows results for $\Gamma = 500$ (dashed
curves) and $\Gamma = 1000$ (solid curves), and for magnetic fields
$B^\prime = 1$\,kG and $B^\prime = 1$\,MG in the upper and lower
panels, respectively.  Note that subtraction of an underlying hard
component will not change the peak $\nu F_\nu$ flux value by more than
$\approx 5$\%.
 
The unattenuated power of the SSC component is comparable to the
synchrotron power when $B^\prime = 1$ kG, whereas the SSC component is
much weaker in the strong-field case. The isotropic jet power for the
$B^\prime = 1$ kG model is $\cong 2.0\times 10^{54}$ erg s$^{-1}$ and
$\cong 5.8\times 10^{53}$ erg s$^{-1}$ for $\Gamma = 500$ and $\Gamma
= 1000$, respectively, and is dominated by the energy in the escaping
radiation. When $B^\prime = 1$ MG, the isotropic jet power is $\cong
1.1\times 10^{55}$ erg s$^{-1}$ and $\cong 7\times 10^{56}$ erg
s$^{-1}$ for $\Gamma = 500$ and $\Gamma = 1000$, respectively, and is
dominated by magnetic-field energy.  For $B^\prime \ll$ 1 kG, the SSC
flux becomes much brighter than the synchrotron flux, and the jet
power becomes dominated by particles, even assuming that all particle
energy is in the form of relativistic electrons.

Because the SSC component is strongly attenuated by $\gamma\gamma$
processes for the $\Gamma$ factors considered, an electromagnetic
cascade will be formed with $\gamma$ rays emerging at lower energies
where the system becomes optically thin.  The timescale for the
electromagnetic radiation to cascade to energy $E_\gamma$ is shorter
than the synchrotron time, given by $t_{\rm syn} \approx 0.006
(\Gamma/1000)^{1/2}/[(B^\prime/{\rm kG})^{3/2} \sqrt{E_\gamma/100{\rm~MeV}}]$
s.  Unless $B^\prime\ll 1$ kG, in which case much more power is found in the
cascading SSC emission than in the synchrotron emission, the cascading
timescale is too short to explain the $\approx 0.2$ s delay between
the GBM and LAT emission.  This model also faces the well-known
line-of-death problem \citep{Preece:98} that the standard synchrotron
mechanism makes a spectrum softer than $\alpha = -2/3$, whereas
$\alpha = -0.48\pm 0.07$ in interval b, representing a nearly $3\sigma$
discrepancy from the hardest expected synchrotron emissivity. For the strong-field case where
the SSC component is weak, the separate hard component in GRB 090510
would then have to originate from a different mechanism.

\subsubsection{Afterglow Synchrotron Model}

\citet{2009MNRAS.400L..75K,2009arXiv0910.5726K} and \citet{2009arXiv0909.0016G,2009arXiv0910.2459G} have
proposed forward shock emission from the early afterglow as the origin
of the delayed onset and the hard component extending into the LAT
energy band. This possibility is also considered by
\citet{2010ApJ...709L.146D}, \citet{2009arXiv0911.4453C}, and \citet{2009ApJ...706L..33G}.
 In particular, \citet{2009arXiv0909.0016G} calculate the
coasting bulk Lorentz factor of the GRB jet by identifying the time of
the peak LAT emission occurring at $\approx$\,\Time{0.7} with the
deceleration time of a relativistic blast wave.  If $\eta$ is the
efficiency to convert bulk kinetic energy to $\gamma$-ray energy,
$\Gamma_0 \approx 2000 ~n^{-1/8} (t_{\rm peak}/0.2~\rm s)^{-3/8}
(\eta/0.2)^{-1/8} (E_{\gamma,\rm iso}/3.5\times 10^{52}~\rm
erg)^{1/8}$, where $n($cm$^{-3}$) is the density of the surrounding medium. This
expression uses an apparent isotropic $\gamma$-ray energy release that
excludes the LAT emission, which if due to synchrotron radiation from
relativistic electrons, would require a radiative efficiency approaching
unity.

Depending on circumburst density, 
the implied Lorentz factor is about 2--4 times the value of $\Gamma_\min$
calculated in the previous section. For this model, the emission
radius corresponding to the time of the peak LAT flux is $R \approx
2.4\times 10^{16} n^{-1/4}$ cm.  At $t\approx t_{\rm peak}$, the
minimum energy electrons in the forward shock radiate synchrotron
photons at energies \\
$h\nu_m \approx
3.6(\epsilon_e/0.1)^2\xi_e^{-2}(\epsilon_B/0.01)^{1/2}E_{53}^{1/2}(t/0.2\,{\rm
s})^{-3/2}\;$MeV assuming an electron injection index of $p \approx
2.4$, where $\xi_e$ is the fraction of the electrons that take part in
the non-thermal power-law component responsible for the observed
emission.

Recent particle in cell
simulations of relativistic collisionless shocks
\citep{2008ApJ...673L..39S,2008ApJ...682L...5S,2009ApJ...695L.189M} 
suggest that $\xi_e$ is fairly small (of the order of a few percent),
which may in our case allow $h\nu_m$ of several MeV even for low
values of $\epsilon_B$. 
For $\epsilon_e = 0.1$, $\epsilon_B = 0.01$, and an isotropic
equivalent kinetic energy of $E_{k,{\rm iso}} = 10^{53}\;$erg (similar
to $E_{\rm\gamma,iso}$), the cooling break frequency is given by
\citep{2002ApJ...568..820G}, 
$h\nu_c \approx 0.4 n^{-1} (\epsilon_e/0.1)^{-1}
(\epsilon_B/0.01)^{-1/2} E_{53}^{-1/2} (t/0.2{\rm~s})^{-1/2}$\,keV.
This expression holds when $\epsilon_B \ll \epsilon_e$ and Klein-Nishina effects
are unimportant, so that
$Y = [(1+4\epsilon_e/\epsilon_B)^{1/2}-1]/2 \approx \sqrt{\epsilon_e/\epsilon_B} > 1$.
This would imply fast cooling $\nu_c \ll \nu_m$ for
$n \sim 1$, which could result in a highly radiative shock for
$\epsilon_e \sim 1$.
However, the late time broad band spectrum at $t \sim 100\;$s from
optical-UV through X-ray and up to the LAT $\gamma$-ray energies
\citep{2010ApJ...709L.146D} suggests $h\nu_c(100\,{\rm s}) \sim 300\;$MeV,
which, even if with a large uncertainty, together with the overall
afterglow modeling suggest a much lower external density of $n \sim
10^{-5}$. Such a low density would imply 
$h\nu_c \approx 0.3 (n/10^{-5})^{-1} (\epsilon_B/0.01)^{-3/2}
E_{53}^{-1/2} (t/0.5{\rm~s})^{-1/2}$\,GeV
(where the
Klein-Nishina effect suppresses SSC cooling), so that $\nu_c$ passes
through the (low part of the) LAT energy range around $\sim 0.5\;$s
from the onset of the main emission episode, or at $\sim T_0+1\;$s,
when there is a softening in the LAT photon index, near the end of the
prompt emission. Thus, after the prompt emission, the LAT energy range
would be above both
$\nu_m$ and $\nu_c$, accounting for the observed photon index. This
would imply 
$h\nu_c \approx 18 (n/10^{-5})^{-1} (\epsilon_B/0.01)^{-3/2}
E_{53}^{-1/2} (t/100{\rm s})^{-1/2}$\,MeV
which is consistent with the broad band spectrum at that time, especially
since the spectral break around $\nu_c$ is very smooth and gradual
\citep{2002ApJ...568..820G}. Interestingly, the inferred value of
$h\nu_m(100\,{\rm s}) \approx 0.43\;$keV and the $\nu_m
\propto t^{-3/2}$ scaling gives $h\nu_m(t_{\rm peak}\approx 0.2\,{\rm s}) 
\approx 4.8\;$MeV, which is very close to the measured value of $\Epeak$ 
near $t_{\rm peak}$. 
This model would not produce a spectrum softer than $\nu F_{\nu} \propto
\nu^{4/3}$, so would have difficulty accounting for the emission near 10\,keV
 in interval b, which appears to be a continuation of the hard spectral
component.

For an adiabatic blast wave and an electron injection index $p$, the
synchrotron flux scales as $\nu F_\nu\propto t^{(2-3p)/4}
\nu^{(2-p)/2}$ at frequencies $\nu > \nu_m,\nu_c$. The measured late 
time LAT flux decay rate of $\nu F_\nu \propto t^{-1.38\pm
0.07}\nu^{-0.1\pm 0.1}$ would in this case require $p = 2.5\pm 0.1$,
while the spectral slope requires $p = 2.2\pm 0.2$. Both are
consistent with $p = 2.4$ at the $1\sigma$ level. This value is also
consistent with the late time afterglow of GRB~090510
\citep{2010ApJ...709L.146D}. 
A radiative blast wave at early times is not required in order to account
for the observed early LAT flux decay rate.


\subsubsection{Hadronic Models}

In hadronic models, photohadronic and proton/ion synchrotron processes
induce electromagnetic cascades, which lead to synchrotron and Compton
emissions from secondary electron-positron pairs
\citep[e.g.,][]{2006NJPh....8..122D, Gupta:07, 2007ApJ...671..645A}.
For a target photon energy distribution $n(\epsilon) \propto
\epsilon^{x}$, the efficiency for photopion processes is $ \propto
R^{-1} \Gamma^{-2} E_{\rm p}^{-1-x} \propto \Gamma^{-4} t_{\rm v}^{-1}
E_{\rm p}^{-1-x}$, where $R\propto c\Gamma^2 t_{\rm v}$
is the shock radius. In this expression, protons with
energy $E_{\rm p}$ preferentially interact with photons with energies
$\propto \Gamma^2/E_{\rm p}$ \citep{1995PhRvL..75..386W, 2009PhRvL.103h1102M}.  The
large deduced values for $\Gamma$ in GRB~090510 make the photopion
efficiency low, so that a very large energy release is required if the LAT
radiation from GRB 090510 is assumed to be from a photomeson-induced
cascade.

A stronger magnetic field shortens the acceleration timescale, leading
to a larger maximum particle energy $E_{\rm max}$, thus enhancing the
photopion production efficiency.  In such a strong magnetic field,
however, the effective injection index of secondary pairs tends to be
about $-2$ \citep{1992MNRAS.258..657C}, which yields a flat spectrum
in a $\nu F_\nu$ plot, while the power-law index of the extra component in
GRB~090510 is $\sim -1.6$.  In a weaker magnetic field, the Compton
component from secondary pairs can harden the spectrum, though with
reduced photopion production efficiency due to the smaller value of
$E_{\rm max}$. The slower cooling time of protons than 
electrons would produce an extended proton-induced emission
feature \citep{Bottcher:98}, though the SSC component would decay even more slowly \citep{Zhang:01}.
A recent numerical calculation by
\citet{2009ApJ...705L.191A} for GRB 090510 demonstrates that the
proton injection isotropic-equivalent power is required to be
larger than $10^{55}$\,erg\,s$^{-1}$ to explain the hard spectra of
the extra component in GRB 090510, which is $\approx 2$ orders of 
magnitude greater than the measured apparent isotropic $\gamma$-ray luminosity.

An alternative hadronic scenario is a proton synchrotron model with a
very strong magnetic field \citep{2009arXiv0908.0513R}.  This model
explains the delayed onset by proton synchrotron radiation in the
prompt phase due to the time required to accelerate, accumulate, and
cool the ultrarelativistic protons.  For an onset times $t_{\rm onset}
\approx 0.2$~s after the start of the GBM main emission at
$T_0+0.5$~s, the required magnetic field is $B^\prime \approx
7.4\times 10^5 ~(\Gamma/1000)^{-1/3} (t_{\rm onset}/0.1~\rm s)^{-2/3}
(E_\gamma/100~\rm MeV)^{-1/3} $~G in the shocked fluid frame.  The
corresponding total energy release, for a two-sided jet beaming factor
$f_b \approx 1.5\times 10^{-4}(\theta_j/1~{\rm deg})^2$, is ${\cal
  E}\approx 1.3\times 10^{54}(\Gamma/1000)^{16/3} (t_{\rm
  onset}/0.1~\rm s)^{5/3} (E_\gamma/100~\rm MeV)^{-2/3}
(\theta_j/1~{\rm deg})^2$~erg \citep[see][for GRB 080916C]{2009ApJ...698L..98W}. 
Thus values of 
$\Gamma \lesssim 1000$ and narrow jet angles $\approx 1^\circ$ are required for
a proton synchrotron scenario, and such strong beaming is not clearly found
in the short hard class of GRBs \citep{2007PhR...442..166N}. 
Proton-dominated GRB models for ultra-high energy cosmic rays
therefore are plausible only with low values of $\Gamma$ 
and narrow jet collimation to reduce the total energy.

\subsection{Implications for the Extragalactic Background Light}

\begin{figure}
  \centering
  \includegraphics[scale=0.55,clip]{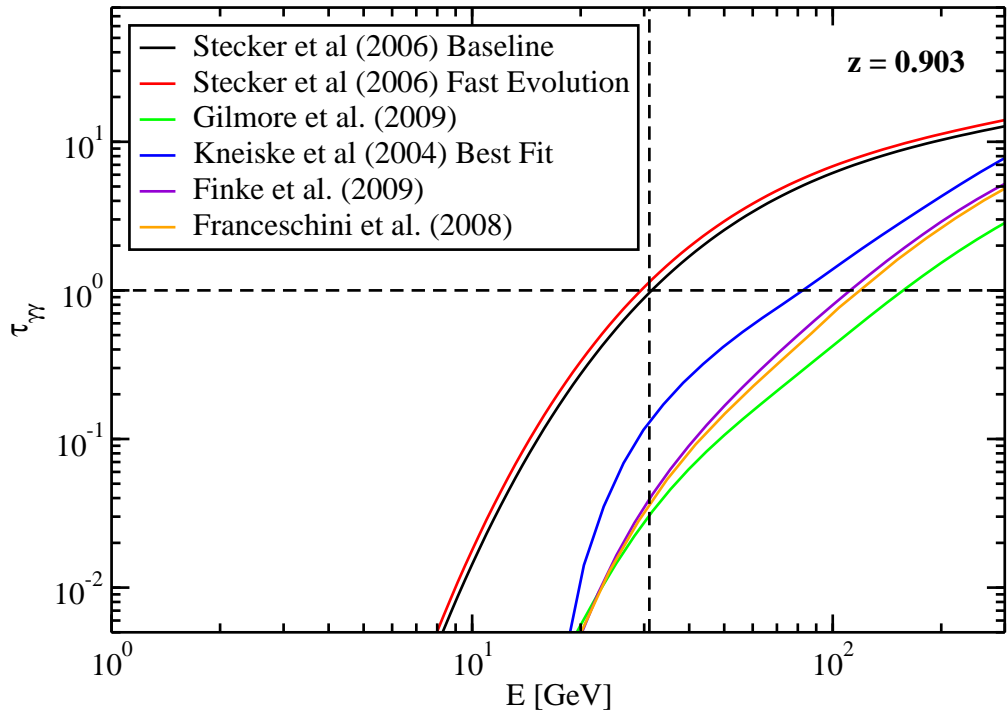}
  \caption{Model predictions of EBL absorption optical depth versus
  photon energy for GRB~090510.  The dashed horizontal line indicates
  an optical depth of unity, and the dashed vertical line shows the
  energy of the highest energy photon.}
  \label{fig:taugg}
\end{figure}

The EBL is dominated by direct starlight in the optical/ultraviolet
and by stellar radiation that is reprocessed by dust in the
infrared. The EBL is difficult to measure directly due to
contamination by zodiacal and Galactic foreground light
\citep{2001ARA&A..39..249H}.  For sources at sufficiently high
redshifts, $\gamma\gamma$ absorption of high-energy $\gamma$-rays by
EBL photons can provide a means of constraining models of the EBL.
Figure~\ref{fig:taugg} shows the absorption optical depth,
$\tau_{\gamma\gamma}$, for various models of the EBL as a function of
$\gamma$-ray energy at the redshift $z=0.903$ of GRB 090510.

We have included curves for the two models of
\citet{2006ApJ...648..774S} as well as the fiducial model of
\citet{2009arXiv0905.1144G}, the best fit model of
\citet{2004A&A...413..807K}, the model by \citet{2008A&A...487..837F},
and ``Model C'' of \citet{2009arXiv0905.1115F}.  The models of
\citet{2006ApJ...648..774S} border on optically thick at 30.5\,GeV; all other
models considered here give a transmission probability of
$e^{-\tau_{\gamma\gamma}}\gtrsim 0.85$.  The baseline and fast
evolution models of \cite{2006ApJ...648..774S} give transmission
probabilities of 0.37 and 0.30, respectively.  Although a higher
energy ($\approx 30.5$\,GeV) photon was found from this burst than for
GRB 080916C ($13$\,GeV; \cite{2009Sci...323.1688A}), that burst was
more constraining for EBL models due to its higher redshift, $z = 4.35
\pm 0.15$ \citep{2009A&A...498...89G}. However, the EBL does evolve with redshift,
and this GRB provides an independent constraint from a later time 
and at a closer distance than GRB 080916C. The low optical depth for the highest
energy photons in GRB~090510 justify neglecting EBL effects in Figure~\ref{fig:dermerSSC}.

\section{Summary}

We have presented \Fermi\ observations of the short hard gamma-ray
burst GRB 090510 during the prompt emission phase extending to 3 s
after the GRB trigger \citep[\Fermi\ and \Swift\ observations at later
times are presented in][]{2010ApJ...709L.146D}.  The apparent
isotropic energy measured by the \Fermi\ GBM and LAT from GRB 090510 
is $(1.08\pm 0.06)\times 10^{53}$\,erg in the energy range from
10\,keV to 30\,GeV, where the upper limit is defined by 
the highest energy photon.  When corrected for corresponding energy
ranges, this is at least
an order-of-magnitude greater than the isotropic energy releases
measured from \Swift\ GRBs with known redshift \citep{2007PhR...442..166N}, 
indicating that GRB 090510 is unusually energetic.  A faint precursor and one LAT
photon with energy $> 100$ MeV are observed $\approx 0.5$ s before
the main pulse, and two other LAT photons are detected before the start of
the bright GBM emission. Large numbers of $> 100$ MeV LAT photons are
detected beginning at $\approx$\,\Time{0.7}.

Spectral analysis shows that the time-integrated flux cannot be fit
with a single Band spectrum, but is well fit with a power-law emission
component in addition to the Band component.  In the time interval
\Time{0.6} to \Time{0.8}, the GRB emission can be resolved into a Band
function with $\alpha = -0.48\pm 0.07$ and a hard spectral power law
with number index $= -1.66$.  The GBM emission is observed to vary on
a timescale of $14\pm 2$ ms in this interval, and on a timescale of
$12\pm2$ ms in the time interval c, from \Time{0.8} to \Time{0.9}. Our
results are summarized in Table~\ref{tab:Gamma_min_results}. In 
time interval c, where the highest energy ($\approx 30.5$\,GeV) photon is detected, 
our most conservative
assumptions give $\Gamma \gtrsim 1000$.

We considered models for the delayed onset of the high-energy LAT
radiation and the appearance of the distinct hard spectral component.
In particular, we considered whether the emission spectrum can be
explained by a synchrotron model for the Band function and SSC
emission for the hard power-law component in the GeV range. Cascading
of SSC emission into the LAT range from higher energies could explain
the hard component, but the delayed onset is found to be too long
compared to the time for development of the cascade. Moreover, the
hard GBM spectrum presents a challenge for synchrotron models that
could instead be explained by a photospheric emission component. Other
models, including external-shock synchrotron and hadronic models, are
also considered to explain the \Fermi\ observations of GRB
090510.  Future \Fermi\ observations of short GRBs will help to
discriminate between models and improve our understanding of
relativistic outflows in GRBs.

\acknowledgments

We thank the anonymous referee for a careful reading of the manuscript
and for many helpful suggestions that have improved the paper.

The \Fermi\ LAT Collaboration acknowledges generous ongoing
support from a number of agencies and institutes that have supported
both the development and the operation of the LAT as well as
scientific data analysis.  These include the National Aeronautics and
Space Administration and the Department of Energy in the United
States, the Commissariat \`a l'Energie Atomique and the Centre
National de la Recherche Scientifique / Institut National de Physique
Nucl\'eaire et de Physique des Particules in France, the Agenzia
Spaziale Italiana and the Istituto Nazionale di Fisica Nucleare in
Italy, the Ministry of Education, Culture, Sports, Science and
Technology (MEXT), High Energy Accelerator Research Organization (KEK)
and Japan Aerospace Exploration Agency (JAXA) in Japan, and the
K.~A.~Wallenberg Foundation, the Swedish Research Council and the
Swedish National Space Board in Sweden.

Additional support for science analysis during the operations phase is
gratefully acknowledged from the Istituto Nazionale di Astrofisica in
Italy and the Centre National d'\'Etudes Spatiales in France.

The \Fermi\ GBM Collaboration acknowledges support for GBM development,
operations, and data analysis from NASA in the US and BMWi/DLR in
Germany.


\clearpage
\begin{deluxetable}{cccccc}
\tablecaption{Durations of GRB 090510 detected with different instruments.}
\tablewidth{0pt}
\tablehead{\colhead{Instrument} &
           \colhead{$T_{90}$ (s)} &
           \colhead{$T_{50}$ (s)} &
           \colhead{Energy Range}}
\startdata
GBM/NaI 3 & 0.6 & 0.2 & 50--300\,keV\\
GBM/NaI 6 & 9.0 & 0.3 & 50--300\,keV\\
GBM/NaI 7 & 1.5 & 0.2 & 50--300\,keV\\
GBM/NaI 3, 6 \& 7 & 2.1 & 0.2 & 50--300\,keV\\
\Swift/BAT & 4.0 & 0.7 & 50--350\,keV\\
INTEGRAL-SPI & 2.5 & 0.1 & 20\,keV--10\,MeV\\
Suzaku-WAM & 5.8 & 0.5 & 50\,keV--5\,MeV\\
\enddata
\label{tab:durations}
\end{deluxetable}

\clearpage
\begin{deluxetable}{clcccccccccccc}
  \tablecaption{$\Gamma_\min$ values for the shortest time scale pulses from
    GRB~090510}
  \tablewidth{0pt}
  \tablehead{\colhead{$T - T_0$} &
    \colhead{Spectrum} &
    \colhead{$t_v$ (ms)} &
    \colhead{$E_\max$ (GeV)} &
    \colhead{$\Gamma_\min$\tablenotemark{a}}\\
	(s) & }
  \startdata
  0.6--0.8 & Band + PL & $14 \pm 2$ & 3.4 & $951 \pm 38$\\
  0.6--0.8 & PL\tablenotemark{b} & $14 \pm 2$ & 3.4 & $703 \pm 34$\\
  0.8--0.9 & Band\tablenotemark{c} & $12 \pm 2$ & 30.5 & $1324 \pm 50$\\
  0.8--0.9 & Band + PL & $12 \pm 2$ & 30.5 & $1218 \pm 61$\\
  0.8--0.9 & PL\tablenotemark{b} & $12 \pm 2$ & 30.5 & $1083 \pm 88$
  \enddata
  \label{tab:Gamma_min_results}
\tablenotetext{a}{Uncertainty on $\Gamma_\min$ is associated with uncertainties in $t_{v}$ and spectral flux only.}
\tablenotetext{b}{Variability time $t_{v}$ for power law (PL) component is assumed to be the same as that
derived from the BGO emission, which is described primarily by Band component. }
\tablenotetext{c}{The Band-function-only fit gives a larger flux than the Band plus power-law 
fit in the portion of the spectrum where the highest-energy $\gamma$ rays typically interact; see Figure~\ref{fig:nuFnu spectra}.  }
\end{deluxetable}

\clearpage
\newcommand\numentry[3]{{$#1^{+#2}_{-#3}$}}
\begin{deluxetable}{clccccccccc}
  \tabletypesize{\scriptsize}
  \rotate
  \tablecaption{Prompt emission spectral fit parameters.}
  \tablewidth{0pt}
  \tablehead{
    \colhead{$T - T_0$} & \colhead{Model} & \multicolumn{4}{c}{Band Model} &
    \multicolumn{3}{c}{power-law or Comptonized} & \colhead{CSTAT / dof} \\
    \cline{3-6} \cline {7-9}\\
    \colhead{(s)} & 
    \colhead{} & 
    \colhead{$A$} &
    \colhead{$\Epeak$} & 
    \colhead{$\alpha$} & 
    \colhead{$\beta$} &
    \colhead{$A$ at 1\,GeV} &
    \colhead{$E_{\rm pk}$} &
    \colhead{Index} & 
    \colhead{}\\
    \colhead{} & \colhead{} & 
    \colhead{($10^{-2}$cm$^{-2}$s$^{-1}$keV$^{-1})$} &
    \colhead{(MeV)} & \colhead{} & \colhead{} &
    \colhead{($10^{-9}$cm$^{-2}$s$^{-1}$keV$^{-1})$} &
    \colhead{(MeV)} & \colhead{} & \colhead{}\\
}
\startdata
0.5--1.0 & Band & \numentry{4.316}{0.116}{0.115} & \numentry{4.104}{0.267}{0.263} & \numentry{-0.75}{0.03}{0.02} & $-2.40 \pm 0.04$ & $\cdots$ & $\cdots$ & $\cdots$ & 1016 / 970\\
 & Band + PL & \numentry{3.188}{0.269}{0.258} & \numentry{3.936}{0.280}{0.260} & \numentry{-0.58}{0.06}{0.05} & \numentry{-2.83}{0.14}{0.20} & \numentry{2.426}{0.531}{0.509} & $\cdots$ & $-1.62\pm 0.03$ & 979 / 968\\
 & Band + Comp & \numentry{3.203}{0.281}{0.266} &\numentry{4.002}{0.285}{0.271} & $-0.59\pm 0.06$ & \numentry{2.94}{0.18}{0.25} & \numentry{3.011}{0.697}{0.658} & \numentry{8.71}{\infty}{4.18} & $-1.60\pm 0.03$ & 976 / 967\\
0.5--0.6 & Band & \numentry{8.047}{0.346}{0.344} & \numentry{2.809}{0.185}{0.174} & $-0.59\pm 0.04$ & $<-5.0$ & $\cdots$ & $\cdots$ & $\cdots$ & 840 / 971\\
0.6--0.8 & Band + PL & \numentry{2.984}{0.365}{0.341} & \numentry{5.102}{0.443}{0.400} & $-0.48 \pm 0.07$ & \numentry{-3.09}{0.21}{0.35} & \numentry{1.862}{0.719}{0.625} & $\cdots$ & $-1.66\pm 0.04$ & 991 / 968 \\
0.8--0.9 & Band & \numentry{0.040}{0.005}{0.004} & \numentry{1.414}{0.928}{0.536} & \numentry{-1.00}{0.11}{0.09} & \numentry{-1.85}{0.05}{0.06} & $\cdots$ & $\cdots$ & $\cdots$ & 886 / 970 \\
 & Band + PL & $0.028 \pm 0.006$ & \numentry{1.894}{1.160}{0.718} & \numentry{-0.86}{0.17}{0.23} & $-3.09$ (fixed) &\numentry{6.439}{1.550}{1.230} & $\cdots$ & \numentry{-1.54}{0.07}{0.04} & 890 / 969\\
0.9--1.0 & PL (LAT only) & $\cdots$ & $\cdots$ & $\cdots$ & $\cdots$ & \numentry{3.721}{1.260}{1.080} & $\cdots$ & \numentry{-1.92}{0.20}{0.22} & 43 / 118\\
\enddata
\label{tab:spectral fits}
\end{deluxetable}

\end{document}